\documentclass{WileyMSP-template}

\usepackage[numbers,sort&compress]{natbib}

\begin{document}

%\pagestyle{fancy}
%\rhead{\includegraphics[width=2.5cm]{vch-logo.png}}

\title{Probing the Nanoscale Excitonic Landscape and Quantum Confinement of Excitons in Gated Monolayer Semiconductors\footnote{This manuscript has been authored by UT-Battelle, LLC, under contract DE-AC05-00OR22725 with the US Department of Energy (DOE). The US government retains and the publisher, by accepting the article for publication, acknowledges that the US government retains a nonexclusive, paid-up, irrevocable, worldwide license to publish or reproduce the published form of this manuscript, or allow others to do so, for US government purposes. DOE will provide public access to these results of federally sponsored research in accordance with the DOE Public Access Plan (http://energy.gov/downloads/doe-public-access-plan).}}

\maketitle

% Author: Please give full first and last names for authors and include * after the name of all corresponding authors

\author{Yueh-Chun Wu*}
\author{Bogdan Dryzhakov}
\author{Huan Zhao}
\author{Ivan Vlassiouk}
\author{Kyle Kelly}
\author{Takashi Taniguchi}
\author{Kenji Watanabe}
\author{Jun Yan}
\author{Benjamin Lawrie*}

% Dedication
\dedication{}

% Affiliations: Please provide adacemic titles (Prof. or Dr.) for all authors where applicable, and include an institutional email address for all corresponding authors
\begin{affiliations}
Yueh-Chun Wu, Benjamin Lawrie\\
Materials Science and Technology Division, Oak Ridge National Laboratory, 1 Bethel Valley Rd, Oak Ridge, TN 37831,USA\\
Email Address: wuy2@ornl.gov, lawriebj@ornl.gov

Bogdan Dryzhakov, Huan Zhao, Ivan Vlassiouk, Kyle Kelly \\
Center for Nanophase Materials Sciences, Oak Ridge National Laboratory, 1 Bethel Valley Rd, Oak Ridge, TN 37831, USA\\

Takashi Taniguchi\\
Research Center for Materials Nanoarchitectonics, National Institute for Materials Science, 1-1 Namiki, Tsukuba 305-0044, Japan\\
Kenji Watanabe\\
Research Center for Electronic and Optical Materials, National Institute for Materials Science, 1-1 Namiki, Tsukuba 305-0044, Japan\\

Jun Yan\\
Department of Physics, University of Massachusetts Amherst, Amherst, Massachusetts 01003, USA
\end{affiliations}

% Keywords: Please provide a minimum of three and a maximum of seven keywords, separated by commas


% Abstract should be written in the present tense and impersonal style (i.e., avoid we), and be at most 200 words long
\begin{abstract}

Engineering and probing excitonic properties at the nanoscale remains a central challenge in quantum photonics and optoelectronics. While exciton confinement via electrical control and strain engineering has been demonstrated in 2D semiconductors, substantial nanoscale heterogeneity limits the scalability of 2D quantum photonic device architectures. In this work, we use cathodoluminescence spectroscopy to probe the excitonic landscape of monolayer $\mathrm{WS_2}$ under electrostatic gating. Exploiting the high spatial resolution of the converged electron beam, we resolve a homojunction arising between gated and ungated regions. Moreover, we reveal an exciton confinement channel arising from an unconventional doping mechanism driven by the interplay between the electron beam and the applied gate fields. These findings offer new insights into the optoelectronic behavior of monolayer semiconductors under the combined influence of electron-beam excitation and electrostatic gating. Our approach provides a pathway for exciton manipulation at the nanoscale and opens opportunities for controlling quantum-confined exciton transport in two-dimensional materials. 

\end{abstract}

% Text: Please use section headings and subheadings as specified below. For communications, all section headings apart from Experimental Section should be removed
% Please make the first reference to a display item bold: \textbf{Figure 1}
% Do not abbreviate Figure, Equation, etc.; display items are always singular, i.e., Figure 1 and 2.
% Equations are always singular, i.e., Equation 1 and 2, and should be inserted using the {equation} environment, not as graphics
% Please do not use footnotes in the text, additional information can be added to the Reference list.

\section{Introduction}
Two-dimensional (2D) transition metal dichalcogenides (TMDs), with their reduced dielectric screening and strong Coulomb interactions, provide a promising platform for realizing a wide range of excitonic many-body states—including superfluids~\cite{liu2022crossover,gupta2020heterobilayers,cutshall2025imaging}, electron-hole liquids~\cite{arp2019electron,qi2023thermodynamic}, and ferromagnetic phases~\cite{wang2022light,ciorciaro2023kinetic}—relevant to quantum simulation and advanced optoelectronic applications. A central challenge in this pursuit is the controlled tailoring of excitonic properties. Confining excitons in reduced dimensions enhances Coulomb interactions and modifies the density of states, unlocking strong exciton–exciton interactions, enabling artificially designed exciton-based logic operations~\cite{high2008control,hu2024quantum}, and paving the way for quantum technologies such as single-photon sources~\cite{aharonovich2016solid,zhao2021site}, scalable quantum simulators~\cite{yu2017moire}, and exciton routers~\cite{liu2020electrically,chen2022chirality}. Recently, techniques have emerged for confining direct excitons through engineered strain profiles and electric fields in 2D semiconductor heterostructures~\cite{luo2023improving,luo2023deterministic,dirnberger2021quasi, heithoff2024valley, thureja2022electrically,hu2024quantum}. 

The characterization of such nanoscale-confined excitons is typically carried out using optical spectroscopy, which offers a high spectral resolution probe of distinct excitonic transitions. However, diffraction-limited far-field optical methods provide only spatially averaged information~\cite{thureja2022electrically,heithoff2024valley,kim2025moire}. While near-field scanning probe techniques offer improved spatial resolution~\cite{kim2024confinement,zhang2022nano,hou2025nanometer}, they often suffer from reduced spectral resolution, particularly when accessing deeply buried or encapsulated material layers required for preserving intrinsic sample quality. Cathodoluminescence (CL) excited by a high-energy converged electron beam, has emerged as a alternative but powerful tool for probing excitonic properties in 2D semiconductors~\cite{luo2023imaging,bonnet2024cathodoluminescence,bonnet2021nanoscale,francaviglia2022optimizing,woo2024engineering,borghi2024cathodoluminescence,ramsden2023nanoscale,zheng2017giant}. Unlike conventional optical techniques, CL utilizes tightly focused electron beams to locally excite carriers, enabling direct examination of excitonic energetics and dynamics at the nanometer scale~\cite{zheng2023electron,xu2024sub,sutter2021cathodoluminescence}.

In this work, we investigate the excitonic emission of monolayer $\mathrm{WS_2}$ under electrostatic gating using CL microscopy, and we resolve the excitonic landscape shaped by the underlying heterostructure stacking. Under electron-beam excitation, we observe gate screening effects, which we attribute to beam-induced trapped carriers in the hBN dielectric. This unconventional electrostatic doping mechanism enables the formation of an exciton confinement potential, giving rise to a localized exciton channel that can be directly visualized via CL microscopy. Our findings elucidate the optoelectronic response of monolayer semiconductors under the combined influence of electron-beam excitation and electrostatic gating. This approach offers a pathway for nanoscale exciton manipulation and opens new opportunities for controlling quantum-confined exciton transport in two-dimensional materials.

\section{Result and Discussion}

\begin{figure}
  \includegraphics[width=\linewidth]{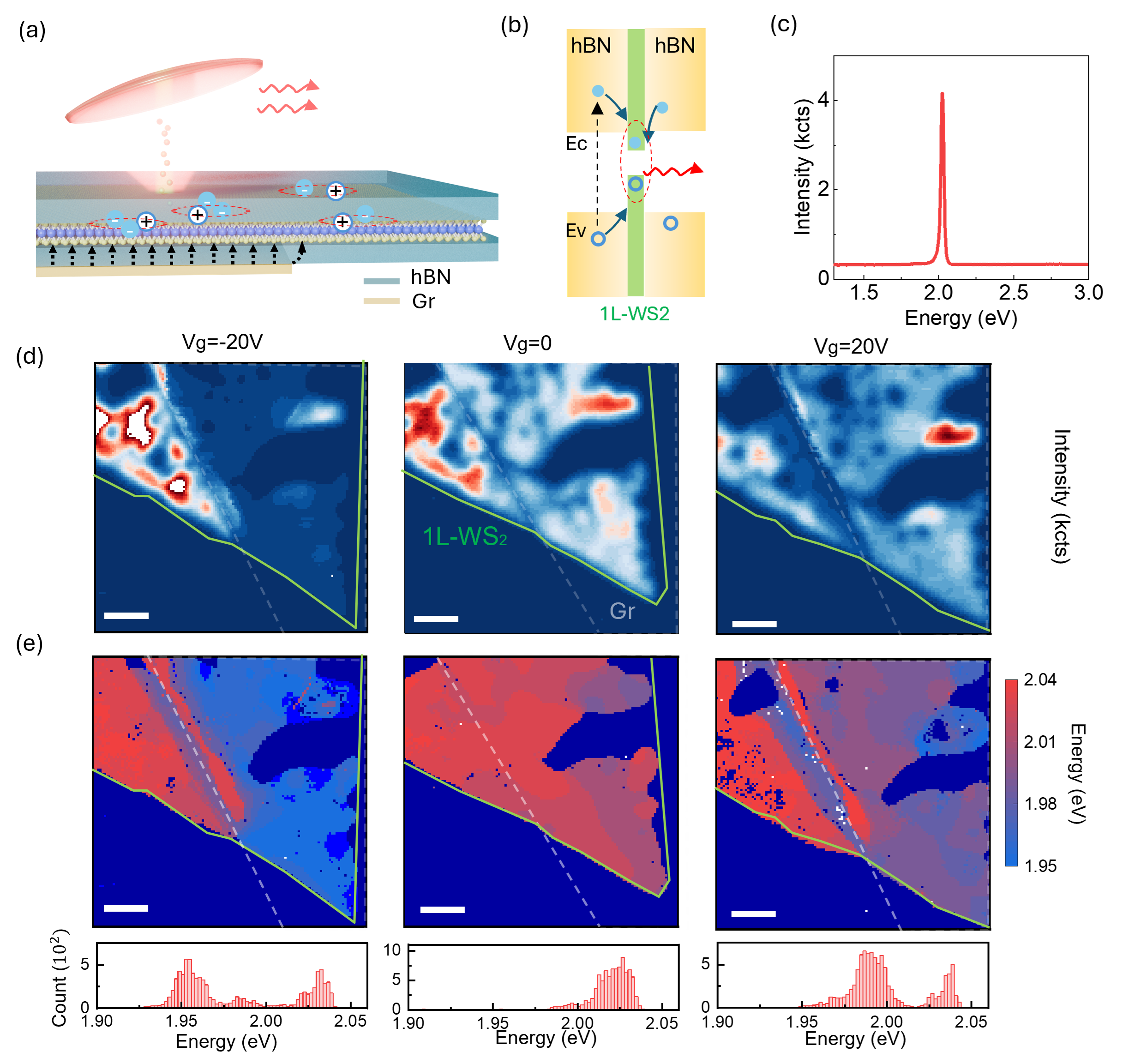}
  \caption{(a) Schematic of CL setup illustrating electron beam excitation of hBN encapsulated, graphene-backgated $\mathrm{WS_2}$ through a parabolic mirror. The CL generated under electron beam excitation is collimated by a parabolic mirror and directed to a spectrometer outside the SEM. (b) CL excitation mechanism in hBN encapsulated $\mathrm{WS_2}$. (c) Prototypical CL spectrum acquired at T=10 K. (d) Intensity maps of T=10 K $\mathrm{WS_2}$ CL integrated over 1.95-2.01 eV at gate voltages of –20V, 0V, and 20V. (e) Exciton emission energy maps at the same gate voltages. Corresponding histograms show the statistical distribution of emission energy peaks extracted from each map. Scale bar: 1$\mathrm{\mu m}$ The green outlines mark the $\mathrm{WS_2}$ flakes, while the white dashed lines indicate the underlying graphene. The entire mapped area is encapsulated by top and bottom hBN layers.}
  \label{fig:Fig1}
\end{figure}

To achieve high sample quality and narrow optical spectral features, both critical for probing intrinsic exciton properties, we encapsulate monolayer $\mathrm{WS_2}$ between hBN layers. Figure~\ref{fig:Fig1}a illustrates the CL measurement setup and device structure, featuring hBN dielectric layers and a few-layer graphene back gate. We note hBN encapsulation is known to significantly enhance CL emission in TMDs~\cite{zheng2017giant,ramsden2023nanoscale,francaviglia2022optimizing}. Without encapsulation, emission from monolayer TMDs is usually too weak to detect due to its low scatting cross-section and potential beam-induced degradation. The interaction volume of the high energy incident electrons leads to carrier generation primarily in the surrounding hBN dielectric. These energetic carriers subsequently relax into the encapsulated $\mathrm{WS_2}$ layer where they recombine radiatively, enhancing the observed CL signal. The underlying CL generation mechanism for the hBN\slash TMD\slash hBN heterostructure is illustrated in Fig.~\ref{fig:Fig1}(b). We note electron-beam exposure can also induce broadband defect CL around 1.9 eV from the underlying $\mathrm{SiO_2}$~\cite{bonnet2024cathodoluminescence,koyama1980cathodoluminescence,curie2022correlative}. To minimize this background emission, we use a 5 keV electron beam at 0.22~nA, ensuring the excitation volume is largely confined within the hBN/1L-$\mathrm{WS_2}$/hBN stack~\cite{ramsden2023nanoscale,francaviglia2022optimizing}, as discussed in the Supporting information, section S1. Under these conditions, the CL spectrum from the encapsulated monolayer exhibits a single, strong emission peak corresponding to 1L-$\mathrm{WS_2}$, as shown in  Fig.~\ref{fig:Fig1}(c).

Figure~\ref{fig:Fig1}(d) displays false-color CL intensity maps acquired at T=10K and spectrally integrated from 1.95eV to 2.01eV, consistent with past observations of emission from monolayer $\mathrm{WS_2}$~\cite{jin2019observation,bonnet2021nanoscale,arora2019excited}. These scans highlight two distinct regions in the vdW stacking: hBN/1L-$\mathrm{WS_2}$/hBN with and without an underlying graphene layer back gate. The excitonic emission intensity is modulated with a gate voltage applied through the underlying graphene. In addition to electrostatic tuning, the nanoscale inhomogeneities in TMD CL like those observed here are known to result from variations in dielectric environment~\cite{bonnet2024cathodoluminescence}, doping~\cite{bonnet2021nanoscale}, strain~\cite{fouchier2019polarized,zheng2025deep}, and interfacial defects~\cite{bonnet2024cathodoluminescence} such as bubbles introduced during the stacking process. These nanoscale inhomogeneities have a direct influence on exciton dynamics, affecting exciton localization, diffusion, recombination pathways, and coupling to external fields. For instance, regions exhibiting suppressed CL emission have been attributed to non-radiative trap states or poor contact at hBN interfaces that inhibit efficient carrier transfer. In additional to the intensity map, Fig.~\ref{fig:Fig1}(e) presents a false-color map of exciton peak energy under various gate voltages. Because of the narrow 15 meV neutral exciton CL linewidth at T=10K, we are able to resolve subtle spectral shifts in exciton energy. At zero gate voltage, the emission peak is centered around 2.03 eV, corresponding to the neutral exciton (X) in monolayer $\mathrm{WS_2}$. Upon gating, the emission redshifts in the gated region, indicative of trion formation.

\begin{figure}
\begin{center}
  \includegraphics[width=\linewidth]{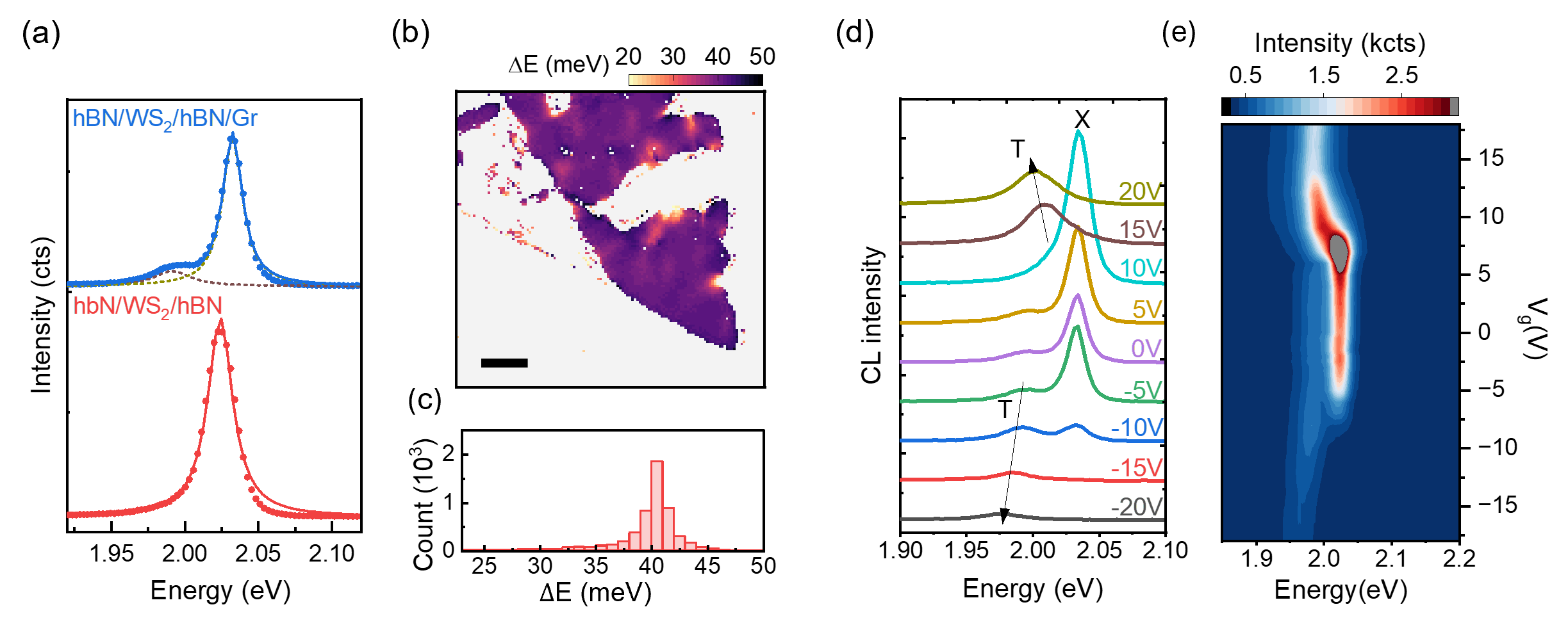}
  \caption{ (a) Representative CL spectra from region with and without underlying graphene at $\mathrm{V_g=0}$. The solid blue line illustrates a double Lorentzian fit, with dashed lines representing the component Lorentzian terms. The red curve is fitted with a single Lorentzian. (b) Map of energy separation $\Delta~E$, given by the difference between the center energies of each component in the double Lorentzian. This map illustrates the trion formation and its binding energy in hBN/$\mathrm{WS_2}$/hBN/Gr at $\mathrm{V_g=0}$. Scale bar: 1 $\mathrm{\mu m}$ (c) Histogram of $\Delta~E$ extracted from (b).  (d) CL spectra at representative gate voltages and (e) gate dependent CL contour map with evolution from neutral exciton (X) to trion (T) with increasing gate voltage amplitude. 
  }
  \label{fig:Fig2}
  \end{center}
\end{figure}

Notably, Fig.~\ref{fig:Fig1}(d) illustrates suppressed emission intensity along the line between the gated and ungated regions at zero gate bias, a point that is explored in greater detail below. The reduced emission intensity may stem from in-plane electric fields promoting exciton dissociation due to doping contrast across the two regions. In addition, tensile strain in the multilayer stack could also contribute to the observed energy shift and intensity modulation~\cite{conley2013bandgap,hernandez2022strain}. This nanoscale modulation of excitonic emission arising from variations in stacking configurations underscores the potential of CL for probing quantum confinement through local strain and electrostatic doping.  

As shown in Fig.~\ref{fig:Fig2}(c), in regions of hBN\slash $\mathrm{WS_2}$\slash hBN without graphene, a single emission peak is centered at 2.024 eV. In contrast, the hBN\slash $\mathrm{WS_2}$\slash hBN\slash Gr region exhibits a blue-shifted exciton peak ($\approx$ 2.032 meV) along with an additional emission feature $\sim$41 meV below the exciton, attributed to a trion. This energy separation is consistent with the previously reported trion binding energy in monolayer $\mathrm{WS_2}$ (30-45meV)~\cite{zheng2025deep,bonnet2021nanoscale,chatterjee2022trion}. In addition, the blue-shifted neutral exciton peak and its reduced linewidth point to exciton-carrier interactions arising from excess free carriers and enhanced screening of impurities~\cite{wu2021enhancement,goldstein2020ground,wu2022negative}. Figures~\ref{fig:Fig2}(d) and (e) show the exciton-trion energy separation obtained from fitting with two Lorentzian peaks across the map. While the exciton emission from the 1L–$\mathrm{WS_2}$ region without underlying graphene is well described by a single peak, the region with graphene clearly exhibits both neutral exciton and trion features. 

We now focus on the gate-dependent excitonic emission of monolayer $\mathrm{WS_2}$ under e-beam excitation. Figures~\ref{fig:Fig2}(d) and ~\ref{fig:Fig2}(e) present the evolution of CL spectra as a function of gate voltage. Electrostatic gating modulates the excitonic landscape in monolayer $\mathrm{WS_2}$ by introducing excess free carriers, which induce oscillator strength transfer from neutral exciton to charged exciton (trion) states~\cite{wu2022negative,arora2019excited}. In the gate-dependent CL spectra, we observe strong neutral exciton emission near $V_g\approx 7$ V, indicating the charge neutrality point. In contrast, trion emission dominates at $V_g < -10$ V and $V_g > 10$ V. The absence of trion emission at smaller gate voltages suggests reduced gating efficiency compared to the expected gating efficiency of $0.8 -1 \times 10^{12} \mathrm{cm^{-2}V^{-1}}$ for hBN layers of 20 - 30 nm thickness determined from a simple capacitance model. Such unconventional gate dependence has been reported in graphene and $\mathrm{MoS_2}$ field effect transistors under electron beam exposure, where e-beam induced charges in the dielectrics screen the gate field and alter the local doping profile~\cite{shi2020reversible}.

\begin{figure}
    \begin{center}
    \includegraphics[width=0.8\linewidth]{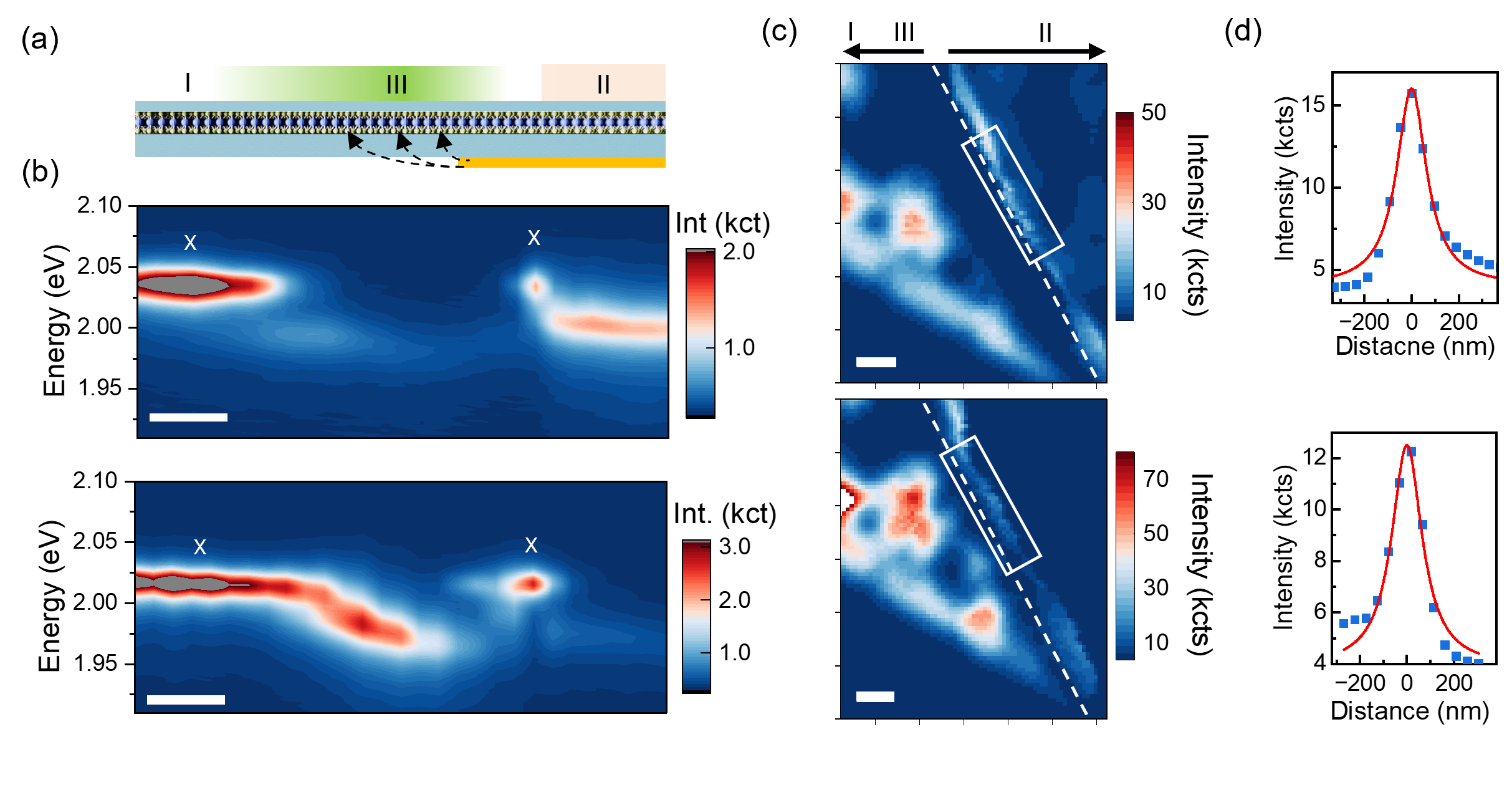}
    \caption{(a) Schematic of the hBN/$\mathrm{WS_2}$/hBN heterostructure with and without back-gate graphene, defining three regions: (I) ungated, (II) gated, and (III) fringing-field region. (b) Contour plots of the measured CL as a function of position across regions I–III at gate voltages of +20 V (top) and –20 V (bottom). Scale bar: 200 nm  (c) Maps of the integrated emission intensity between 2.00–2.067 eV, highlighting the neutral exciton (X) emission at +20 V (top) and –20 V (bottom). The white dashed line outlines the edge of underlying graphene. Scale bar: 500 nm. (d) Spatial profile of the CL-observed confinement channel of X with gate bias +20V (top) and -20V (bottom), obtained by averaging the signal across the channel, as marked by the white solid rectangular area. All spectra were acquired at $T = 10$ K.
    }
    \label{fig:Fig3}
    \end{center}
\end{figure}

With a finite back gate geometry, the electrostatic gating creates doping contrast between $\mathrm{hBN/WS_2/hBN}$ and $\mathrm{hBN/WS_2/hBN/Gr}$ regions, giving rise to a homojunction in the vdW stacking. Indeed, the data shown in Fig.~\ref{fig:Fig1}(e) with $V_g=$ 20 V and -20 V, illustrates a non-trivial excitonic emission profile varying spatially across the two regions. In Figure~\ref{fig:Fig3}(b), representative CL emission contour maps across the junction with $V_g$= 20 V (above) and $V_g$=-20 V (below) highlight three distinct regions: (I) intrinsic $\mathrm{WS_2}$ with minimal gate-induced doping, where neutral excitons dominate, (II) gate-doped $\mathrm{WS_2}$ with prominent trion emission, and (III) a fringing field region, where electric fields emanating from the edge of the gate electrode induce a non-uniform doping profile, as illustrated in Fig.~\ref{fig:Fig3}(a). Remarkably, we observe the emergence of a narrow neutral exciton channel between region II and region III.  This spatially localized channel, situated between the gated and fringing regions (Fig. 3(c)), suggests the formation of an n-i-p (p-i-n) junction, resembling the doping profiles engineered via asymmetric dual gate structures~\cite{thureja2022electrically}. 

\begin{figure}[bth]
\begin{center}
  \includegraphics[width=\linewidth]{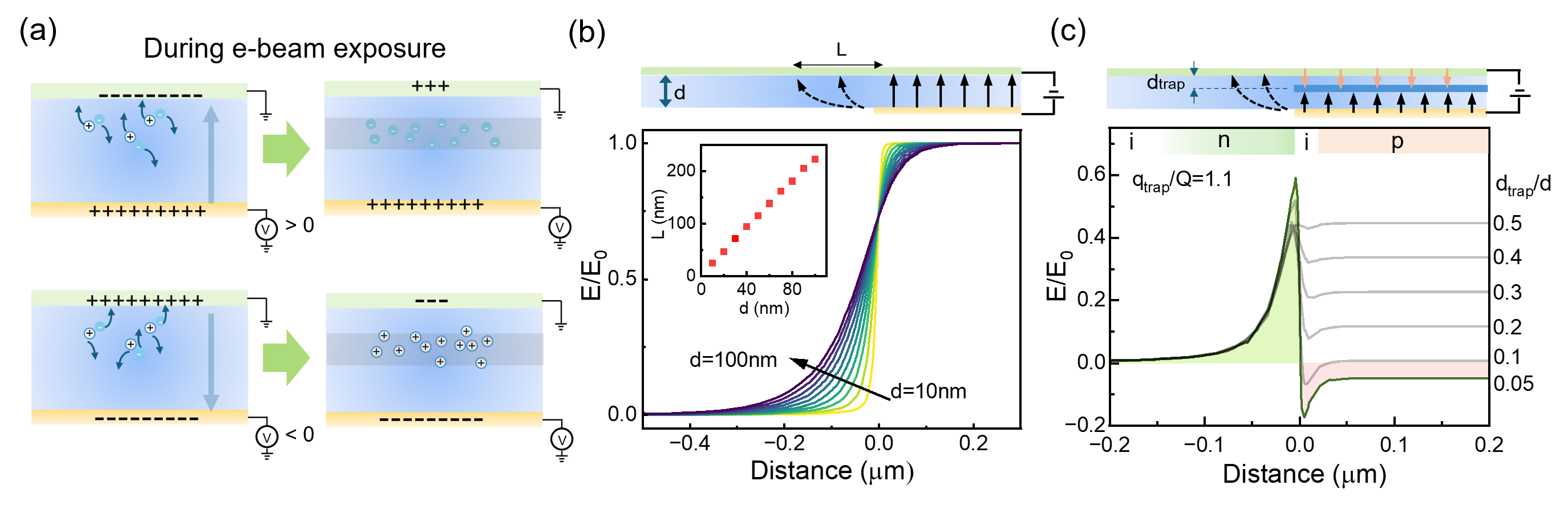}
  \caption{(a) Electron-beam–induced doping under gate bias fields (top: $\mathrm{V > 0}$, bottom: $\mathrm{V < 0}$). Hot electron–hole pairs generated by the electron beam drift across the hBN dielectric under the applied gate bias. The resulting charge trapping in the hBN modifies the electrostatic doping of the monolayer $\mathrm{WS_2}$ (ground plane). (b) Fringing-field simulation of a finite-sized gate electrode coupled to a ground plane through a dielectric ($\epsilon_r$ = 3.9) of varying thickness. (Inset) Fringing-field characteristic length $L$ as a function of plate distance $d$ from the ground plane. (c) Proposed mechanism for neutral exciton confinement, arising from the competition between electrostatic fields from the gate electrode and electron-beam–induced charges in the hBN dielectric. $q_\mathrm{trap}$ denotes the trapped charge density in the hBN dielectric, and $Q$ represents the accumulated charge with the conventional two-plate capacitor model. The effective trapping-charge layer distance $d_\mathrm{trap}$ is compared with the dielectric thickness ($d$=25 nm). The resulting contrast in the electrostatic field profile with $\frac{d_\mathrm{trap}}{d}=0.05$ and 0.1 gives rise to a local p–i–n junction near the edge of the finite gate electrode. 
  }
  \label{fig:Fig4}
  \end{center}
\end{figure}

We note that with the TMD layer being grounded, such a spatially contrasting excitonic emission with and without underlying graphene back gate can not be simply explained with e-beam induced charging to the vdW device. Instead, these observed gate dependent CL spectra and maps indicate an unconventional electrostatic doping mechanism under electron beam excitation. We adopt the previously proposed doping mechanism~\cite{shi2020reversible} that primary doping effects under e-beam exposure are from hot carrier generation, drift under gate bias field, and relaxation to trap states in the dielectric. A schematic of such e-beam-induced doping through hot carrier generation in the dielectric with applied gate voltage is shown in Fig.~\ref{fig:Fig4}(a). This trapped charge layer introduces a screening effect that alters the effective gating behavior. Notably, the emergence of a charge-neutral junction near the edge of the gate electrode further supports the presence of such trapped charges, whose fringing fields can compete with those originating from the gate electrode itself.

To investigate the influence of fringing fields from a finite-sized gate, we use the Ansys Maxwell 2D electrostatic solver. The system is modeled as a finite metallic plate held at a fixed potential, coupled to an effectively infinite ground plane through a dielectric ($\epsilon \approx 3.9$). Our simulations reveal that electric fields emanating from the gate edge extend laterally to the ground plane over a characteristic distance $L$, depending on the gate-to-ground separation $d$. These fringing fields induce a spatially varying doping profile near the gate edge. With the fringing field profile depending on the plate separation, e-beam induced charges in the dielectric can lead to a reverse doping profile near the edge of the gate electrode. Figure 4(b) shows the electric field profile at the ground plane as a function of normalized trapping layer depth $d_{trap}/d$, where $d_{trap}$ is the distance of the charge layer from the ground plane. The trapping charge density is modeled by the ratio $\frac{q_{trap}}{Q}$, where Q corresponds to the charge density in the conventional two-plate model. When $q_{trap}$ shares the same polarity as Q, the gate field is effectively screened. Moreover, a reverse doping profile emerges across the region II and III when $q_{trap}>Q$ and results in the formation of lateral n-i-p and p-i-n junctions. To explore this effect in detail, we simulate various combinations of trapping charge density ($q_{trap}/Q$) and normalized trapping depth, as discussed in the Supporting Information.  A representative case with $\frac{q_{trap}}{Q} \approx 1.1$ and  $\frac{d_{trap}}{d}$ ranging from 0.05 to 0.5, shown in Fig. 4(b), captures the experimentally observed doping distribution, delineating distinct regions: intrinsic, gated, fringing field, and importantly, the emergence of a charge neutral zone at the interface between the electrostatic p-doped and n-doped region.

\section{Conclusion}
In conclusion, we have demonstrated gate-dependent CL spectroscopy of monolayer $\mathrm{WS_2}$ encapsulated in hBN, revealing nanoscale electrostatic doping behavior under electron beam excitation. The spatially resolved CL measurements reveal the nanoscale excitonic landscape and the formation of locally confined neutral excitons due to the fringing field effect. Through electrostatic modeling and simulations, we attribute these unconventional electrostatic gating effects to charge trapping within the hBN dielectric. We highlight that this unconventional electrostatic doping under electron beam exposure offers a promising strategy for engineering quantum potentials for excitons in vdW heterostructures. Such engineered potentials open new avenues for controlling exciton-exciton interaction and provide a platform for exploring excitonic many-body states. Indeed, this could offer a critical additional degree of nanoscale control that is otherwise challenging to achieve in van der Waals quantum simulators~\cite{kennes2021moire}. Importantly, this work presents a direct visualization of exciton behavior at a nanoscale homojunction, made possible by the high spatial resolution of CL and the high sample quality afforded by hBN encapsulation. The CL-resolved charge neutrality junctions exhibit widths of approximately 150 nm. We note that the spatial resolution of CL in hBN/$\mathrm{WS_2}$/hBN stacks is limited by carrier diffusion within the hBN dielectric layer~\cite{francaviglia2022optimizing}. The resolution can be further improved by using thinner hBN, albeit at the cost of reduced CL intensity, and by lowering the temperature to suppress carrier diffusion in hBN. These trade-offs require optimization for future studies of excitonic quantum phenomena at the nanoscale via cathodoluminescence.

% Experimental section

\section{Experimental Section}
\threesubsection{Device Fabrication}\\
The atomic flakes of $\mathrm{WS_2}$, hexagonal boron nitride (hBN), and few-layer graphene were first exfoliated on Si wafers with 300 nm of $\mathrm{SiO_2}$ and inspected under an optical microscope. To make high quality 1L-$\mathrm{WS_2}$ heterostructures, we annealed hBN in an $\mathrm{O_2}$/Ar atmosphere (50 sccm /200 sccm) at 500$^{\circ}$ C for 2--3 h before stacking. Using a dry transfer technique with a polypropylene carbonate (PPC) stamp, the flakes were then stacked with a few-layer graphene flake as the back gate and transferred to substrates with premade electrodes. All the exfoliation, inspection, and stacking processes were completed in a glovebox with controlled humidity and oxygen both $<$ 0.01 ppm to minimize sample degradation. The stacked samples were thermally
annealed at 350$^{\circ}$ C for 1 h in an argon environment to improve the quality.

The heterostructure morphology was measured using atomic force microscopy (AFM). The top hBN layer is approximately 160 nm thick, while the bottom hBN layer is about 23 nm thick. Few layer graphene, serving as the backgate, is estimated to be 2-3 nm ($\approx$ 5--10 layers) thick based on the optical contrast. 

\threesubsection{CL Measurements}\\
Cathodoluminescence (CL) measurements were performed in an FEI Quattro environmental SEM integrated with a Delmic Sparc CL collection module, which uses a parabolic mirror to collect emission under electron-beam excitation. CL spectra were acquired with a 5 kV, 220 pA electron beam and an exposure time of 1 s, using a 150 l/mm grating and a 100 µm input slit. All measurements were carried out at a temperature of T = 10 K.

\threesubsection{Simulation}\\
The simulations were performed with the Ansys Maxwell 2D electrostatic solver. To model the fringing electric fields that emerge from the edge of a plate coupled to a ground plane, we simulate two plates of defined thickness ($t$) and width ($W$), separated by a dielectric layer of thickness ($d$) and relative permittivity $\epsilon_r \approx 3.9$. The gate plate, which is half the size of the ground plate, is biased at 1 V. We set the dimensions so that $W \gg d \gg t$ and focus our analysis on the fringing fields at the edge of the gate plate. The detailed setup and results with various trapping charge density and depth can be found in supporting information.

\medskip
\textbf{Supporting Information} \par %Please delete the Suppporting Information statement if it is not applicable. Please supply Supporting Information in another file. Supporting information should not be provided in .tex format
Supporting Information is available from the Wiley Online Library or from the author.

% Acknowledgements
\medskip
\textbf{Acknowledgements} \par %delete if not applicable))
This research was sponsored by the U. S. Department of Energy, Office of Science, Basic Energy Sciences, Materials Sciences and Engineering Division. CL microscopy was performed through a user project supported by the Center for Nanophase Materials Sciences (CNMS), which is a US Department of Energy, Office of Science User Facility at Oak Ridge National Laboratory. J.Y. acknowledges support from the National Science Foundation (DMR-2004474) for device fabrication. K.W. and T.T. acknowledge support from the JSPS KAKENHI (Grant Numbers 21H05233 and 23H02052), the CREST (JPMJCR24A5), JST and World Premier International Research Center Initiative (WPI), MEXT, Japan for hBN synthesis.

% References
\medskip

% Use the following code if you wish to generate your bibliography with BibTeX;
% replace the string "MSP-template" below with the name(s) of
% the BibTeX data base(s) you want to use.
% The resulting bibliography-output (the content of the .bbl file)
% must be pasted back into this file before submission.
% Please also include your BibTeX data base file(s) in your submission
% so that we can re-run BibTeX if necessary.
%
%\bibliographystyle{MSP}
%\bibliography{sample}

% Figures/tables and captions
% Permission statements are required for all figures reproduced or adapted from previously published articles/sources. Please also ensure that all necessary permissions to reproduce images have been received
% Please remove these statements for original figures

% Table of contents entry should be 50 - 60 words long
% Image should be 55 mm broad and 50 mm high or 110 mm broad and 20 mm high

\begin{figure}
\textbf{Table of Contents}\\
\medskip
  \includegraphics{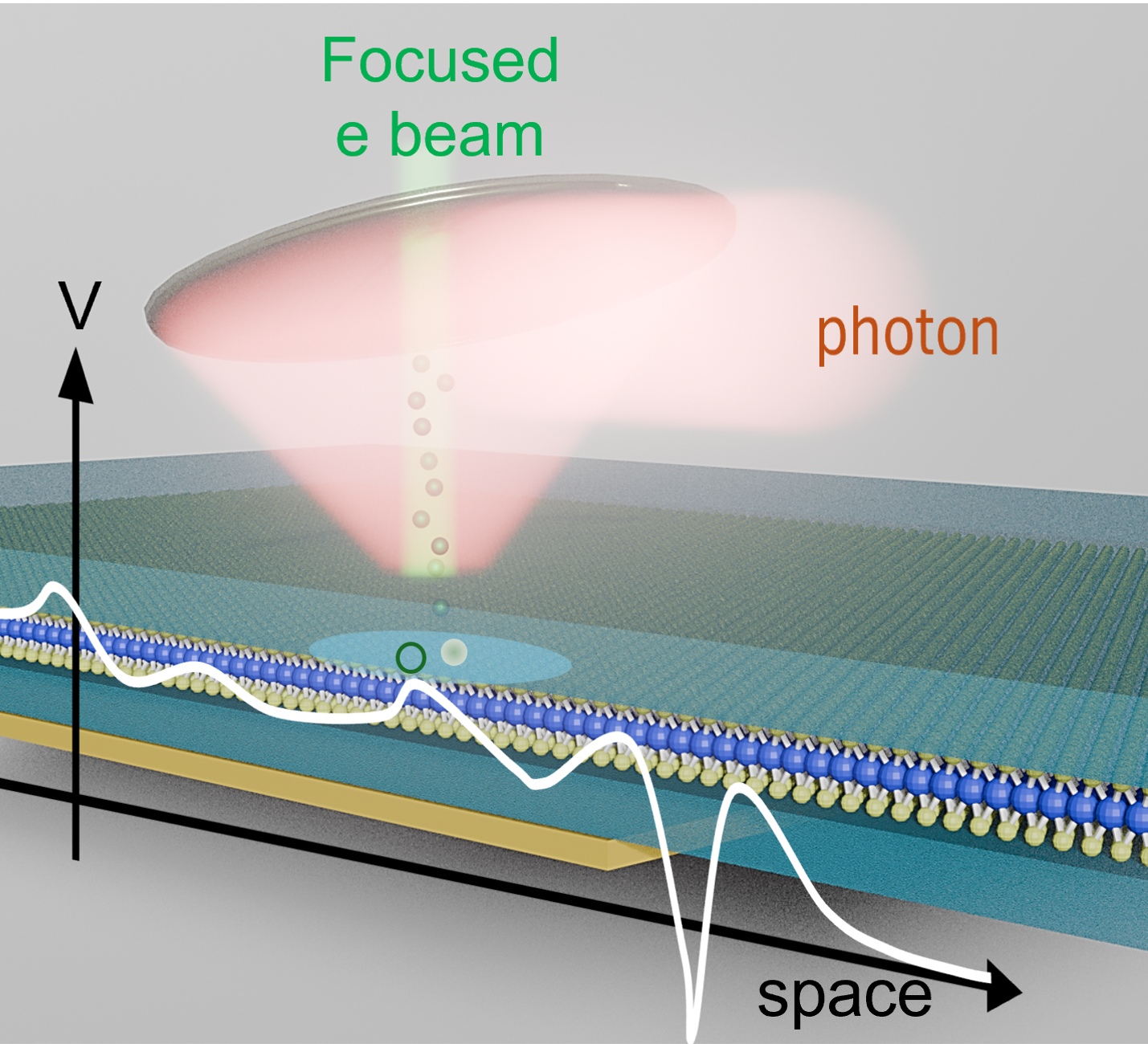}
  \medskip
  \caption*{Nanoscale resolution of excitonic landscape and confinement potential in gated monolayer semiconductors using cathodoluminescence}
\end{figure}

\end{document}